%% file: bindings.tex
\documentclass{llncs}
\usepackage{amsmath,amssymb}
\usepackage{stmaryrd}
\usepackage{tikz}
\usetikzlibrary{shapes,calc}
\usepackage{url}
\usepackage{multirow}
\usepackage{multicol}
\usepackage{rotating}
\usepackage{listings}
\usepackage{lmodern} 
\usepackage{mathptmx} 
\usepackage[utf8]{inputenc}
\usepackage[T1]{fontenc}
\DeclareSymbolFont{letters}{OML}{txmi}{m}{it}
\DeclareMathAlphabet{\mathcal}{OMS}{cmsy}{m}{n}
\usepackage{bussproofs}
\usepackage{hyphenat}
\usepackage{enumitem} 
\usepackage{fancyvrb}
\usepackage{listings}

\makeatletter
\g@addto@macro \normalsize {%
 \setlength\abovedisplayskip{2.2pt}%
 \setlength\belowdisplayskip{2.2pt}%
}

\newenvironment{myitem}[1][]
  {\itemize[leftmargin=*,topsep=0.3ex,itemsep=1pt, #1]}
  {\enditemize}

\makeatother

\let\llncssubparagraph\subparagraph
\let\subparagraph\paragraph
\usepackage[compact]{titlesec}
\let\subparagraph\llncssubparagraph
\usepackage{titlesec}
\titlespacing*{\section}{0.5pt}{1.1\baselineskip}{0.3\baselineskip}
\titlespacing*{\subsection}{0.5pt}{0.65\baselineskip}{0.25\baselineskip}

\usepackage{color}

\definecolor{light-gray}{gray}{0.85}
\newcommand\coll[1]{\mbox{\colorbox{light-gray}{$\!#1\!$}}}

\usepackage{amsthm} 

\usepackage{varioref}

\usepackage[
   a4paper,
   pdftex,  
   pdfkeywords={},
   pdfborder={0 0 0},
   draft=false,
   bookmarksnumbered,
   bookmarks,
   bookmarksdepth=2,
   bookmarksopenlevel=2,
   bookmarksopen]{hyperref}

\usepackage{cleveref}
\usepackage{cite}
\usepackage{calc}

\include{myCommands}

\allowdisplaybreaks

\allowdisplaybreaks
\usepackage{thmtools}
\usepackage{thm-restate}
\usepackage{enumitem}

\makeatletter
\def\thm@space@setup{%
  \thm@preskip=1.22ex
  \thm@postskip=\thm@preskip 
}
\makeatother

\begin{document}

\title{A Formalized General Theory of Syntax with Bindings\thanks{This technical report is an extended version of the conference paper \cite{our-own-paper}. }
}
\author{Lorenzo Gheri\inst{1} \and Andrei Popescu\inst{1,2}}
\institute{
Department of Computer Science, 
Middlesex University London, UK
\and 
Institute of Mathematics Simion Stoilow of the Romanian Academy, Bucharest, Romania
}

\maketitle

\begin{abstract}
\vspace*{-5ex}
We present the formalization of a theory of syntax with bindings that has 
been developed and refined over the last decade to support several large formalization 
efforts. Terms are defined for an arbitrary number of constructors of varying 
numbers of inputs, quotiented to alpha-equivalence and sorted 
according to a binding signature. 
The theory includes a rich collection of properties of the standard operators  
on terms, such as substitution and freshness. 
It also includes induction and recursion principles and support for semantic interpretation, 
all tailored for smooth interaction with the bindings and the standard operators.  
\end{abstract}

\vspace*{-5ex}
\section{Introduction}

Syntax with bindings is an essential ingredient in the formal specification and 
implementation of logics and programming languages. However, correctly and formally specifying, 
assigning semantics to,
and reasoning about bindings is notoriously difficult and error-prone. This fact is 
widely recognized in the formal verification community 
and is reflected in 
manifestos and benchmarks 
such as the influential POPLmark challenge \cite{POPLmark}. 
 
In the past decade, in a framework developed intermittently 
starting with the second author's PhD \cite{pop-thesis} 
and moving into the first author's ongoing PhD, 
a series of results in logic and $\lambda$-calculus have been formalized in 
Isabelle/HOL \cite{nipkow-et-al-2002,nipkow-klein-2014}.  
These include classic results (e.g., FOL completeness and soundness of 
Skolemization \cite{blanchette-et-al-2014-ijcar,soundCompl-jou,blanchette-frocos2013}, 
$\lambda$-calculus standardization and Church-Rosser theorems \cite{pop-recPrin,pop-thesis}, System F 
strong normalization \cite{pop-HOASOnFOAS}), as well as 
the meta-theory of Isabelle's Sledgehammer 
tool \cite{blanchette-frocos2013,blanchette-et-al-2013-types}.  
 
In this paper, we present the Isabelle/HOL formalization of the framework itself (made available from
the paper's website~\cite{binding-scripts}). 
While concrete system syntaxes 
differ in their 
details, there are some fundamental phenomena 
concerning bindings 
that follow the same generic  
principles. It is these fundamental phenomena that our framework aims to capture, 
by mechanizing a form of universal algebra for bindings. The framework has evolved 
over the years 
through feedback from 
concrete 
application challenges: Each time a tedious, seemingly routine construction was encountered, 
a question arose as to whether this could be performed once and for 
all in a syntax-agnostic fashion. 

The paper is structured as follows. We start with an example-driven overview of our design decisions 
(Section~\ref{sec-exa}). Then we present the general theory: 
terms as alpha-equivalence classes of ``quasiterms,'' standard operators on terms 
and their basic properties (Section~\ref{GenSet}), 
custom induction and recursion schemes (Section~\ref{sec-reas}), 
including support for the semantic interpretation of syntax, and the sorting 
of terms according to a signature (Section~\ref{sec-sorting}). 
%
Within the large body of formalizations in the area (Section~\ref{sec-RelWork}), 
distinguishing features of our work are the general setting (many-sorted signature, possibly infinitary syntax), 
a rich theory of the standard operators, and operator-aware recursion.

\section
{Design Decisions}
\label{sec-exa}

In this section, we use some examples to motivate our design choices for the theory. 
We also introduce conventions and notations that will be relevant throughout the paper.

The paradigmatic example of syntax with bindings is 
that of the $\lambda$-calculus \cite{bar-lam}. 
We assume an infinite supply of variables, $x \in \var$. The $\lambda$-terms, 
$X,Y \in \term_\lambda$, are defined by the following BNF grammar:
$$
\begin{array}{lll}
X    &::=& \Var\;x \mid \App\;X\;Y  \mid \Lm\;x\;X
\end{array}
$$
Thus, a $\lambda$-term is either 
a variable, or an application, or a $\lambda$-abstraction. 
This grammar specification, while sufficient for first-order abstract syntax, 
is incomplete when it comes to syntax with bindings---%
we also 
need to indicate which 
operators introduce bindings and in which of their arguments. 
Here, $\Lm$ is the only binding operator: When applied to the variable 
$x$ and the term $X$, it binds $x$ in $X$. 
After knowing the binders, the usual convention is to {\em identify terms modulo alpha-equivalence}, 
i.e., to treat as equal terms that only differ in the names of 
bound variables, 
such as, e.g., 
$\Lm\,\coll{x}\,(\App\;(\Var\,\coll{x})\;(\Var\;y))$ and 
$\Lm\,\coll{z}\,(\App\;(\Var\,\coll{z})\;(\Var\;y))$. 
The end results of our theory will involve terms modulo alpha.
We will call the raw terms 
``quasiterms,'' reserving the word ``term'' for alpha-equivalence classes.     

\subsection{Standalone Abstractions}
\label{prel-abs}

To make the binding structure manifest, we will ``quarantine'' the bindings and their associated 
intricacies into 
the 
notion of {\em abstraction}, which 
is a pairing of a variable and a term, again modulo alpha. For example, for the $\lambda$-calculus we will have
$$
\begin{array}{cc}
X    \;::=\; \Var\;x \mid \App\;X\;Y  \mid \Lam\;A
\hspace*{6ex}&\hspace*{6ex}
A    \;::=\; \Abs\;x\;X
\end{array}
$$
where $X$ are terms and $A$ 
abstractions. Within $\Abs\;x\;X$, we assume that 
$x$ is bound in $X$. 
%
The $\lambda$-abstractions $\Lm\;x\;X$ of the the original syntax     
are now written $\Lam\;(\Abs\;x\;X)$.   
%

\subsection{Freshness and Substitution}
\label{prel-freshSubst}

The two most fundamental and most standard operators on $\lambda$-terms are: 
\begin{myitem}
\item the freshness predicate, $\fresh : \var \ra \term_\lambda \ra \bool$, 
where $\fresh\;x\;X$ states that $x$ is fresh for (i.e., does not occur free in) $X$; 
for example, it holds that $\fresh\;x\;(\Lam\;(\Abs \allowbreak\;x\;(\Var\;x)))$ and $\fresh\;x\;(\Var\;y)$ (when $x\not=y$), 
but not that $\fresh\;x\;(\Var\;x)$. 
\item the substitution operator, $\_[\_/\_] : \term_\lambda \ra \term_\lambda \ra \var \ra \term_\lambda$, 
where $Y\,[X/x]$ denotes the (capture-free) substitution of 
term $X$ for (all free occurrences of) 
variable $x$ in 
term $Y$; 
e.g., 
if $Y$ is $\Lam\;(\Abs\;x\;(\App\;(\Var\;x)\;(\Var\;y)))$ and $x \not\in \{y,z\}$, then: 
\begin{myitem}
\item $Y\,[(\Var\;z)/y] = \Lam\;(\Abs\;x\;(\App\;(\Var\;x)\;(\Var\;z)))$
\item $Y\,[(\Var\;z)/x] = Y$ (since bound occurrences like those of $x$ in $Y$ are not affected)
\end{myitem}
\end{myitem}
And there are corresponding operators for abstractions---e.g., $\freshAbs\;x\;(\Abs\;x\;(\Var\;x))$ holds. 
Freshness and substitution are pervasive in the meta-theory of $\lambda$-calculus, as well as in most logical 
systems and formal semantics of programming languages. The basic properties of these operators 
lay at the core of 
important meta-theoretic results in these fields---our formalized theory aims at the exhaustive 
coverage of these basic properties.  

\subsection{Advantages and Obligations from Working with Terms Modulo Alpha}

In our theory, we start with defining quasiterms and \abstractions{} and their alpha-equivalence. 
Then, after proving all the syntactic constructors and standard operators to be compatible with alpha, we quotient 
to alpha, obtaining what we call terms and abstractions, and define the versions of these operators on 
quotiented items. 
%
For example, 
let $\qterm_\lambda$ and $\qabs_\lambda$ be the types  
of quasiterms and \abstractions{} in $\lambda$-calculus. Here, the \abstraction{} constructor, 
$\qAbs : \var \ra \qterm_\lambda \ra \qabs_\lambda$, 
is a free 
constructor, of the kind produced by standard datatype 
specifications \cite{berghofer-wenzel-1999,blanchette-et-al-2014-tru}. 
%
The types $\term_\lambda$ and $\abs_\lambda$ are $\qterm_\lambda$ and $\qabs_\lambda$ quotiented to alpha.  
We prove compatibility of $\qAbs$ with alpha 
and then define $\Abs : \var \ra \term_\lambda \ra \abs_\lambda$ by lifting $\qAbs$ to quotients. 

The decisive advantages of working 
with quasiterms and \abstractions{} modulo alpha, i.e., with terms and abstractions, 
are that 
(1) substitution behaves well (e.g., is compositional) 
and 
(2) Barendregt's variable convention \cite{bar-lam} (of assuming, w.l.o.g.,   
the bound variables fresh for the parameters) can be invoked in proofs. 

%
However, this choice brings the obligation to prove that all concepts on terms 
are compatible with alpha. Without employing suitable abstractions, 
this can become quite difficult 
even in the most ``banal'' contexts. 
Due to nonfreeness, primitive recursion 
on terms requires a proof that the definition is well formed, i.e., that the overlapping 
cases lead to the same result. 
%
As for Barendregt's convention, its rigorous usage in 
proofs needs a principle that goes 
beyond the usual structural induction for free datatypes. 

A framework that deals gracefully with these obligations can make an important 
difference in applications---enabling the formalizer to quickly leave behind low-level ``bootstrapping'' issues 
and move to the interesting core of the results. 
To address these obligations, we formalize 
state-of-the-art techniques from the literature \cite{pitts-AlphaStructural,UrbanTasson,pop-recPrin}.

%

\subsection{Many-Sortedness} \label{prel-manySorted}

While $\lambda$-calculus has only one syntactic category of terms (to which we 
added that of abstractions for convenience), this is often not the case. 
FOL has two: 
terms and formulas. The Edinburgh Logical Framework (LF) \cite{har-fra} has three: 
object families, type families and kinds. More complex calculi 
can have many syntactic categories. 

Our framework will 
capture these phenomena. 
We will call the syntactic categories {\em sorts}. 
We will distinguish syntactic categories for terms (the sorts) from those for variables (the {\em varsorts}). 
Indeed, e.g., 
in FOL we do not have variables ranging over formulas,  
in the $\pi$-calculus \cite{MilPiBook} we have channel names 
but no process variables, etc.  

Sortedness is important, but formally quite heavy. 
In our formalization, we postpone dealing with 
it for as long as possible. 
We introduce an 
intermediate notion of {\em good} term, for which we are able to 
build the bulk of the theory---only as the very last step we introduce many-sorted signatures
and transit from ``good'' to ``sorted.''

\subsection{Possibly Infinite Branching} \label{prel-infBranch}

Nominal Logic's \cite{pitts01nominal,UrbanTasson} 
notion of finite support 
has become central in state-of-the-art techniques for reasoning about bindings. 
Occasionally, however, important developments 
step outside finite support. For example, 
(a simplified) 
CCS \cite{MilCCSBook} has the following syntactic categories 
of data expressions $E \in \expp$ and processes $P \in \procc$: 
$$
\begin{array}{cc}
E    \;::=\; \Var\;x \mid 0 \mid E + E
\hspace*{6ex}
&
\hspace*{6ex}
P    \;::=\; \Inp\;c\;x\;P \mid \Out\;c\;e\;P \mid 
\sum_{i\in I}P_i 
\end{array}   
$$
Above, $\Inp\;c\;x\;P$, usually written $c(x).\,P$, is an input prefix $c(x)$ followed by a 
continuation process $P$, with $c$ being a channel 
and $x$ a variable which is bound in $P$. Dually, 
$\Out\;c\;E\;P$, usually written $c\,\ov{E}.\,P$, is an output-prefixed process
with $E$ an expression. 
The exotic constructor here is the sum $\sum$, which models nondeterministic choice from a collection $(P_i)_{i \in I}$ of 
alternatives indexed by a set $I$. It is important that $I$ is allowed to be infinite, 
for modeling 
different decisions based on different received inputs. 
But then process terms may use infinitely many variables, i.e., may not be finitely supported.  
Similar issues 
arise 
in infinitary FOL \cite{kei-mod} and Hennessey-Milner logic \cite{henessy-milner-logic}. 
In our theory, we cover such 
infinitely branching syntaxes.  

\section
{General Terms with Bindings}
\label{GenSet}

We start the presentation of our formalized theory, in its journey 
from quasiterms (\ref{subsec-qterms}) to terms via alpha-equivalence (\ref{subsec-alpha}). 
The journey is fueled by the availability of fresh variables, 
ensured by cardinality assumptions on constructor branching 
 and variables (\ref{subsec-good}).   
%
It culminates with a systematic   
study of the standard term operators 
(\ref{subsec-termsTh}).

\subsection{Quasiterms}
\label{subsec-qterms}

The types $\qterm$ and $\qabs$, of quasiterms and \abstractions{}, are defined as mutually recursive datatypes 
polymorphic in the following type variables: $\indexx$ and $\bindexx$, of indexes for 
free and bound arguments, $\varsort$, of varsorts, i.e., sorts of variables, 
and $\opsym$, of (constructor) operation symbols. 
For readability, below we omit the occurrences of these type variables as parameters to $\qterm$ and $\qabs$: 
$$
\begin{array}{rrcl}
\textsf{datatype } &
  \qterm &\;=\;& \qVar\;\varsort\;\var  \mid 
            \\
         &&&
   \qOp\;\opsym\;((\indexx,\qterm)\,\inputt)\;((\bindexx,\qabs)\,\inputt)
\\
\textsf{and } &
  \qabs &\;=\;& \qAbs\;\varsort\;\var\;\qterm
\end{array}
$$
\par
Thus, any quasiabstraction has the form $\qAbs\;\xs\;x\;X$, putting together the variable $x$ 
of varsort $\xs$ 
with the quasiterm $X$, indicating the binding of $x$ in $X$. 
On the other hand, 
a quasiterm is either an injection $\qVar\;\xs\;x$, of a variable $x$ of varsort $\xs$, 
or has the form $\qOp\;\delta\;\inp\;\binp$ , i.e., consists of 
an operation symbol 
applied to 
some inputs that can be either free, $\inp$, or bound, $\binp$. 
%

We use $(\alpha,\beta)\,\inputt$ as a type synonym for $\alpha \ra \beta\;\option$, 
the type of partial functions from $\alpha$ to $\beta$; such a function returns either 
$\None$ (representing ``undefined'') or $\Some\;b$ for $b : \beta$. This type models 
inputs to the quasiterm constructors of varying number of arguments. An operation symbol 
$\delta : \opsym$ can be applied, via $\qOp$, to:
(1) a varying number of free inputs, i.e., 
families of quasiterms modeled as members of $(\indexx,\qterm)\,\inputt$
and 
(2) a varying number of bound inputs, i.e., 
families of \abstractions{} modeled as members of $(\indexx,\qabs)\,\inputt$.
%
%
For example, taking $\indexx$ to be $\nat$ we capture $n$-ary operations 
for any 
$n$ (passing to $\qOp\;\delta$ inputs   
defined only on $\{0,\ldots,n-1\}$), as well as as countably-infinitary operations 
(passing to $\qOp\;\delta$ inputs defined on the whole $\nat$). 

Note that, so far, we consider sorts of variables but not sorts of terms. 
The latter will come much later, in Section~\ref{sec-sorting}, when we introduce signatures. 
Then, we will gain 
control 
(1) on which varsorts 
should be embedded in which term sorts and 
(2) on which operation symbols are allowed to be applied to which sorts 
of terms. 
But, until then, we will develop the interesting part of the theory of bindings without sorting the terms.

On quasiterms, we define 
freshness, $\qFresh : \varsort \ra \var \ra \qterm \ra \bool$, 
substitution, $\_[\_ / \_]_{\_} : \qterm \ra \qterm \ra \var \ra \varsort \ra \qterm$, 
parallel substitution, $\_[\_] : \qterm \ra (\varsort \ra \var \ra \qterm\;\option) \ra \qterm$, 
swapping, $\_[\_ \wedge \_]_{\_} : \qterm \allowbreak\ra \var \ra \var \ra \varsort \ra \qterm$, and alpha-equivalence, 
$\al : \qterm \ra \qterm \ra \bool$---and corresponding operators on \abstractions{}: $\qFreshAbs$, 
$\alAbs$, etc. 

The definitions proceed as expected, with picking suitable fresh 
variables in the case of substitutions and alpha. For parallel substitution, 
given a 
(partial) variable-to-quasiterm assignment $\rho: \varsort \ra \var \ra \qterm\;\option$, 
the quasiterm $X[\rho]$ is obtained by substituting, 
for each free variable $x$ of sort $\xs$ in $X$ for which $\rho$ is defined, the quasiterm $Y$ where  
$\rho\;\xs\;x = \Some\;Y$. 
%
We only show the formal definition of alpha.

\subsection{Alpha-Equivalence}
\label{subsec-alpha}

\begin{figure}[t]
\hspace*{-1ex}
{\footnotesize
$
\begin{array}{rcl}
\al\;(\qVar\;\xs\;x)\;(\qVar\;\xs'\;x') &\iff&
   \xs = \xs' \;\wedge\; x = x'
\\
\al\,(\qOp\;\delta\;\inp\;\binp)\,(\qOp\;\delta'\;\inp'\;\binp') &\;\iff\;&
   \delta = \delta' \,\wedge\, \lift\,\al\;\inp\;\inp' \,\wedge\, \lift\,\alAbs\;\binp\;\binp' 
\\
\al\;(\qVar\;\xs\;x)\;(\qOp\;\delta'\;\inp'\;\binp') &\iff&
   \Ffalse
\\
\al\;(\qOp\;\delta\;\inp\;\binp)\;(\qVar\;\xs'\;x') &\iff&
   \Ffalse
\\
\alAbs\;(\qAbs\;\xs\;x\;X)\;(\qAbs\;\xs'\;x'\;X')  &\iff&
    \xs = \xs' \;\wedge\;(\exists y \notin \{x,x'\}.\;\qFresh\;\xs\;y\;X \;\wedge\;
\\
&&
 \qFresh\;\xs\;y\;X' \;\wedge\; \al\;(X [ y \wedge x]_{xs})\;(X' [ y \wedge x']_{xs}))
\end{array}
$
}
\vspace*{-2ex} 
\caption{Alpha-Equivalence} 
\label{fig-alpha}
\vspace*{-4ex}
\end{figure}

We define the 
predicates $\al$ (on quasiterms) and $\alAbs$ (on \abstractions{}) 
mutually recursively, as shown in Fig.~\ref{fig-alpha}.  For variable quasiterms, we require 
equality on both the variables and their sorts. For $\qOp$ quasiterms, we 
recurse through the components, $\inp$ and $\binp$. Given any predicate $P : \beta^2 \ra \bool$, we write 
$\lift\,P$ for its lifting to $(\alpha,\beta)\,\inputt^2 \ra \bool$, 
defined as $\lift\,P\;\inp\;\inp' \iff \forall i.\;\mbox{ case }(\inp\;i,\inp'\;i) \mbox{ of } 
(\None,\None) \allowbreak \Ra \Ttrue \mid(\Some\;b,\,\Some\;b') \Ra P\;b\;b' \mid\_ \Ra \Ffalse$.
Thus, $\lift\,P$ relates two inputs  
just in case they have the same domain and their 
results are componentwise 
related.

\begin{conv}\rm
Throughout this paper, we write $\lift$ for the 
natural 
lifting of the various operators from terms and abstractions to free or bound inputs. 
\end{conv}

In Fig.~\ref{fig-alpha}'s clause for \abstractions{}, we require that the bound variables 
are of the same sort and 
there exists some fresh $y$ such that $\al$ holds 
for the terms where $y$ is swapped with the bound variable. Following 
Nominal Logic, 
we prefer to use swapping instead of substitution in alpha-equivalence, 
since this 
leads to simpler proofs.

\subsection{Good Quasiterms and Regularity of Variables}
\label{subsec-good}

In general, 
$\al$ will not be an equivalence, namely, 
will not be transitive: Due to the arbitrarily wide branching of the constructors, 
we may not always have fresh variables $y$ 
available in an attempt to prove transitivity by induction. 
To remedy this, we restrict ourselves to ``good'' quasiterms, whose constructors do not branch  
beyond the cardinality of $\var$. Goodness is defined as the 
mutually recursive predicates 
$\qGood$ and $\qGoodAbs$: 
$$
\begin{array}{rcl}
\qGood\;(\qVar\;\xs\;x) &\iff&
   \Ttrue
\\
\qGood\;(\qOp\;\delta\;\inp\;\binp)  &\;\iff\;&
   \lift\,\qGood\;\inp \;\wedge\; \lift\,\qGoodAbs\;\binp \;\wedge\;
\\ &&
|\dom\;\inp| < |\var| \;\wedge\; |\dom\;\binp| < |\var|
\\
\qGoodAbs\;(\qAbs\;\xs\;x\;X) &\iff&
    \qGood\;X
\end{array}
$$
where, given a partial function $f$, we write $\dom\;f$ for its domain. 
%


Thus, for good items, we hope to always have a supply of fresh variables. Namely, we
hope to prove 
$
\qGood\;X \Lra \forall \xs.\;\exists x.\;\qFresh\;\xs\;x\;X
$. 
But goodness is not enough. 
We also need a special property for the type $\var$ of variables. 
In the case of finitary syntax, it suffices to take $\var$ to be 
countably infinite, since a finitely branching term will contain fewer than $|\var|$ variables 
(here, meaning a finite number of them)---this can be proved by induction on terms, 
using 
the fact that a finite union of finite sets is finite. 

So let us attempt to prove the same in our general case. 
In the inductive $\qOp$ case, we know from goodness 
that the branching is smaller than $|\var|$, so to conclude we would need the following: 
{\em A union of sets smaller than $|\var|$ indexed by a set smaller than $|\var|$ 
stays smaller than $|\var|$.} It turns out that this is a well-studied property 
of cardinals, called {\em regularity}---with $|\nat|$ being the smallest regular cardinal. 
Thus, the desirable generalization of countability is regularity 
(which is available from Isabelle's cardinal library \cite{cardHOL}). 
Henceforth, we will assume: 

\begin{ass}\rm \label{ass-reg}
$|\var|$ is a regular cardinal. 
\end{ass}

We will thus have not 
only one, but a $|\var|$ number of fresh variables: 

\begin{prop}\rm
$\qGood\;X \Lra \forall \xs.\;|\{x.\;\qFresh\;\xs\;x\;X\}| = |\var|$
\end{prop}

Now we can prove, for good items, the properties of alpha familiar from 
the $\lambda$-calculus, including it being an equivalence and an alternative formulation 
of the abstraction case, where ``there exists a fresh $y$'' is replaced with 
``for all fresh $y$.'' While the ``exists'' variant is useful when proving 
that two terms are alpha-equivalent, the ``forall'' variant gives stronger inversion and induction 
rules for proving implications from $\al$. 
(Such fruitful ``exsist-fresh/forall-fresh,'' or ``some-any'' dychotomies have been previously discussed in the context of bindings, e.g, in \cite{DBLP:conf/tphol/NorrishV07,aydemirPOPL08,MillTiu-proofThGenJudg}.)

\begin{prop}\rm \label{lem-alpha}
The following hold: 
\\(1) $\al$ and $\alAbs$ are equivalences on good quasiterms and \abstractions{}
\\(2) The predicates defined by replacing, in Fig.~\ref{fig-alpha}'s definition, the abstraction 
case with  
$$
\begin{array}{c}
\alAbs\;(\qAbs\;\xs\;x\;X)\;(\qAbs\;\xs'\;x'\;X')  \;\iff
\\
    \xs = \xs' \wedge (\coll{\forall}\! y \notin \{x,x'\}.\, \qFresh\,\xs\,y\,X \wedge
 \qFresh\,\xs\,y\,X' \!\coll{\Lra}\! \al (X [ y \wedge x]_{xs}) (X' [ y \wedge x']_{xs})) 
\end{array}
$$
coincide with $\al$ and $\alAbs$. 
\end{prop}

\subsection{Terms and Their Properties}\label{subsec-termsTh}

We define $\term$ and $\abs$ as collections 
of $\al$- and $\alAbs$- equivalence classes of $\qterm$ and $\qabs$. 
Since $\qGood$ and $\qGoodAbs$ are compatible with $\al$ and $\alAbs$, we 
lift them 
to corresponding predicates on terms and abstractions, $\good$ and $\goodAbs$. 
%
\leftOut{
It is routine to show that, through the canonical projections 
$\proj : \qterm \ra \term$ and $\projAbs : \qabs \ra \abs$, good items correspond to good quasiitems:

\begin{prop}\rm \label{lem-good}
For all quasiterms $X$ and \abstractions{} $A$, it holds that 
$\qGood\;X \iff \good (\proj\;X)$ and $\qGoodAbs\;A \iff \goodAbs\;(\projAbs\;A)$
\end{prop}
}


We also prove that all constructors and operators 
are alpha-compatible,   
which allows lifting them to terms: 
$\Var : \varsort \ra \var \ra \term$, 
$\Op : \opsym \ra (\indexx,\term)\allowbreak\inputt \ra (\bindexx,\abs)\,\inputt \ra \term$, 
$\Abs : \varsort \ra \var \ra \term \ra \abs$, 
$\fresh : \varsort \ra \term \ra \bool$, $\_[\_ / \_]_{\_} : \term \ra \term \ra \var \ra \varsort \ra \term$, etc. 

\leftOut{
\begin{myitem}
\item The syntactic constructors:
\begin{myitem} 
\item embedding of variables, $\Var : \varsort \ra \var \ra \term$
\item application of operation symbol, $\Op : \opsym \ra (\indexx,\term)\inputt \ra (\bindexx,\abs)\inputt \ra \term$
\item constructor of abstractions, $\Abs : \varsort \ra \var \ra \term \ra \abs$
\end{myitem}
\item The standard operators: 
\begin{myitem}
\item freshness, $\fresh : \varsort \ra \term \ra \bool$, 
\item substitution, $\_[\_ / \_]_{\_} : \term \ra \term \ra \var \ra \varsort \ra \term$, 
\item swapping, $\_[\_ \wedge \_]_{\_} : \term \ra \var \ra \var \ra \varsort \ra \term$
\end{myitem}
\end{myitem}

Note that, even though the terms are not sorted, the operators contain varsort 
information. For example, $Y[X/x]_{xs}$ will substitute, in $Y$, $X$ for all occurrences of 
$x$ with sort $x$, i.e., as $\Var\;\xs\;x$. We need this information in order to 
the operators to behave as expected later, when we sort the terms. 
} 

To establish an abstraction barrier that sets terms free from their quasiterm origin, 
we prove that the syntactic constructors 
mostly behave 
like 
free constructors, in that $\Var$, $\Op$ and $\Abs$ are exhaustive and 
$\Var$ and $\Op$ are injective and nonoverlapping. 
True to the quarantine principle expressed in Section~\ref{prel-abs},  
the only nonfreeness incident occurs for 
$\Abs$. 
Its equality behavior 
is regulated by the ``exists fresh'' and ``forall fresh'' 
properties inferred from the definition of $\alAbs$ and Prop.~\ref{lem-alpha}(2), respectively: 

\begin{prop}\rm \label{lem-Abs}
Assume $\good\;X$ and $\good\;X'$. Then the following are equivalent:
\\(1) $\Abs\;\xs\;x\;X = \Abs\;\xs'\;x'\;X'$ 
\\(2) $\xs = \xs' \,\wedge\,(\exists y \notin \{x,x'\}.\;\fresh\;\xs\;y\;X \,\wedge\,
 \fresh\;\xs\;y\;X' \,\wedge\, X [ y \wedge x]_{xs} = X' [ y \wedge x']_{xs})$
\\(3) $\xs = \xs' \,\wedge\,(\forall y \notin \{x,x'\}.\;\fresh\;\xs\;y\;X \,\wedge\,
 \fresh\;\xs\;y\;X' \Lra X [ y \wedge x]_{xs} = X' [ y \wedge x']_{xs})$
\end{prop}

Useful rules for abstraction equality also hold with 
substitution:

\begin{prop}\rm \label{lem-Abs-subst}
Assume $\good\;X$ and $\good\;X'$. Then the following hold: 
\\(1) $y \notin \{x,x'\} \,\wedge\, \fresh\;\xs\;y\;X \,\wedge\,
 \fresh\;\xs\;y\;X' \,\wedge\, X\,[(\Var\;\xs\;y)\,/\,x]_{xs} = X'\,[(\Var\;\xs\;y)\,/\,x']_{xs} \Lra \Abs\;\xs\;x\;X = \Abs\;\xs\;x'\;X'$ 
\\(2) $\fresh\;\xs\;y\;X \;\Lra\; \Abs\;\xs\;x\;X = \Abs\;\xs\;y\;(X\,[(\Var\;\xs\;y)\,/\,x]_{xs})$
\end{prop}

To completely seal the abstraction barrier, for all the standard operators we prove 
simplification rules regarding their interaction with the constructors, 
which makes the former behave as if they had been defined in terms of the latter. 
For example, the following facts resemble an inductive definition of freshness (as a predicate): 

\begin{prop}\label{lem-imp-fresh}\rm
Assume $\good\;X$, 
$\lift\,\good\;\inp$, $\lift\,\good\;\binp$, 
$|\dom\;\inp| < |\var|$ and $|\dom\;\binp| < |\var|$. 
The following hold:
\\(1) $(\ys,y) \not= (\xs,x) \,\Lra\, \fresh\;\ys\;y\;(\Var\;\xs\;x)$
\\(2) $\lift\,(\fresh\;\ys\;y)\;\inp \,\wedge\, \lift\,(\freshAbs\;\ys\;y)\;\binp \;\Lra\; 
\fresh\;\ys\;y\;(\Op\;\delta\;\inp\;\binp)$
\\(3) $(\ys,y) = (\xs,x)\,\vee\, \fresh\;\ys\;y\;X \;\Lra\; \freshAbs\;\ys\;y\;(\Abs\;\xs\;x\;X)$
\end{prop}

Here and elsewhere, when dealing with $\Op$, we make cardinality 
assumptions on the domains of the inputs to make sure the  
terms $\Op\;\delta\;\inp\;\binp$ are good. 

We can further improve on Prop.~\ref{lem-imp-fresh}, obtaining ``iff'' facts that resemble a primitively recursive 
definition of freshness (as a function): 

\begin{prop}\label{lem-simp-fresh}\rm
Prop.~\ref{lem-imp-fresh} stays true if the implications 
are replaced by equivalences ($\!\iff\!$). 
\end{prop}

For substitution, we prove facts with a similarly primitive recursion flavor:

\begin{prop}\label{lem-simp-subst}\rm
Assume $\good\;X$, $\good\;Y$, $\lift\,\good\;\inp$, $\lift\,\good\;\binp$, 
$|\dom\;\inp| < |\var|$ and $|\dom\;\binp| \allowbreak< |\var|$. 
The following hold:
\\(1) $(\Var\;\xs\;x)\,[Y/y]_{ys} \;=\; (\mbox{if $(\xs,x) = (\ys,y)$ then $Y$ else $\Var\;\xs\;x$})$ 
\\(2) $(\Op\;\delta\;\inp\;\binp)\,[Y/y]_{ys} = 
\Op\;\delta\;(\lift\,(\_[Y/y]_{ys})\,\inp)\;(\lift\,(\_[Y/y]_{ys})\,\binp)$ 
%
%
\\(3) $(\xs,x) \not= (ys,y) \;\wedge\; \fresh\;\xs\;x\;Y
\;\Lra\; 
(\Abs\;\xs\;x\;X)\,[Y/y]_{ys} \,=\, \Abs\;\xs\;x\;(X\,[Y/y]_{ys})
$
\end{prop}
 
We also prove generalizations of Prop.\ \ref{lem-simp-subst}'s facts for parallel substitution, for example, 
$
\lift\,(\fresh\;\xs\;x)\;\rho \allowbreak \Lra 
(\Abs\;\xs\;x\;X)\,[\rho] \,=\, \Abs\;\xs\;x\;(X\,[\rho])
$. 
%

Note that, for properties involving $\Abs$, the simplification rules 
require 
freshness of the bound variable: 
$\freshAbs\;\ys\;y\;(\Abs\;\xs\;x\;X)$ is reducible to $\fresh\;\ys\;y\;X$ 
only if $(\xs,x)$ is distinct from $(\ys,y)$, 
$(\Abs\;\xs\;x\;X)\,[Y/y]_{ys}$ is expressible in terms of $X\,[Y/y]_{ys}$ 
only if $(\xs,x)$ is distinct from $(\ys,y)$ and fresh for $Y$, etc. 

Finally, we 
prove lemmas that regulate the interaction between the standard operators, in all possible combinations: 
freshness versus swapping, freshness versus substitution, substitution versus substitution, etc. 
Here are a few samples: 

\begin{prop}\label{lem-long}\rm
If the terms $X,Y,Y_1,Y_2,Z$ are $\good$ and the assignments $\rho,\rho'$ are $\lift\,\good$,   
%
then: 
\\(1) Swapping distributes over all operators, including, e.g., substitution:
$$Y\,[X/x]_{xs}\,[z_1 \wedge z_2]_{zs} \;=\; (Y\,[z_1 \wedge z_2]_{zs})\,[(X[z_1 \wedge z_2]_{zs})\,/\,(x[z_1 \wedge z_2]_{xs,zs}) ]_{xs}$$
where $x[z_1 \wedge z_2]_{xs,zs} \;=\; (\mbox{if $\xs = \zs$ then $x[z_1 \wedge z_2]$ else $x$})$
\\(2) Substitution of the same variable (and of the same varsort) distributes over itself: 
$$
X\ [Y_1 / y]_{ys}\, [Y_2 / y]_{ys} \;=\; X\,[(Y_1\,[Y_2 / y]_{ys}) / y]_{ys}
$$
\noindent
(3) Substitution of different variables distributes over itself, assuming 
freshness: 
$$(\ys \neq \zs\ \lor\ y \neq z) \,\wedge\, \fresh\;\ys\;y\;Z
\;\Lra\; 
X\,[Y / y]_{ys}\, [Z / z]_{zs} \,=\, (X\,[Z / z]_{zs})\ [(Y\,[Z / z]_{zs}) / y]_{ys}
$$
\noindent
(4) Freshness for a substitution decomposes into freshness for its participants: 
$$\fresh\;\zs\;z\;(X[Y / y]_{ys}) 
\iff
((\zs,z)=(\ys,y) \lor \fresh\;\zs\;z\;X) \,\land\,  (\fresh\;\ys\;y\;X \lor \fresh\;\zs\;z\;Y)$$
\noindent
\leftOut{
(5) Parallel substitution distributes over unary substitution:
$$
X\,[Y / y]_{ys}\,[\rho] \,=\, X\,[\rho\,[y \la Y[\rho]]_{ys}]
$$
where $\rho[y \la Y[\rho]]$ is the assignment $\rho$ updated with value $\Some\,(Y[\rho])$ for $y$. 
\\
} 
(5) Parallel substitution is compositional: 
$$X\,[\rho]\,[\rho'] \;=\; X\,[\rho \bullet \rho']$$
where $\rho \bullet \rho'$ is the monadic composition of $\rho$ and $\rho'$, defined as 
$$
(\rho \bullet \rho')\,\xs\;x \,=\; \mbox{case $\rho\;\xs\;x$ of $\None$ $\Ra$ $\rho'\;\xs\;x$ $\mid$ $\Some\;X$ $\Ra$ $X[\rho']$}
$$
\end{prop}

In summary, we have formalized quite exhaustively the general-purpose properties of all syntactic constructors 
and standard operators. 
%
Some of these properties are 
subtle. In formalization of concrete results for particular syntaxes, 
they are likely to require a lot of time to even formulate them correctly, let alone prove them---which would be 
wasteful, 
since they are independent on the particular syntax.

\section{ Reasoning and Definition 
Principles} \label{sec-reas} 

We formalize schemes for induction (\ref{Ind}), recursion and semantic interpretation (\ref{sec-RecDef}) that 
realize the Barendregt 
convention and are compatible 
with the standard operators. 

\subsection{Fresh Induction} \label{Ind}

We introduce fresh induction by an example. 
To prove Prop.~\ref{lem-long}(4), we use (mutual) structural induction over terms and abstractions, 
proving the statement together with the corresponding statement for abstractions,  
$\freshAbs\;\zs\;z\;(A[Y / y]_{ys}) 
\iff
((\zs,z) = (\ys,y) \lor \freshAbs\;\zs\;z\;A) \,\land\,  (\freshAbs\;\ys\;y\;A \lor \fresh\;\zs\;z\;Y)
$. 
The proof's only interesting case 
is the $\Abs$ case, say, for abstractions of the form 
$\Abs\;\xs\;x\;X$. 
However, 
if we were able to assume freshness of $(\xs,x)$ for all the statement's parameters, 
namely $Y$, $(\ys,y)$ and $(\zs,z)$,  
this case would also become ``uninteresting,'' following automatically from the induction hypothesis 
by mere 
simplification, as shown below (with the freshness assumptions highlighted):  
%
$$
\begin{array}{l}
\freshAbs\;\zs\;z\;((\Abs\;\xs\;x\;X)\,[Y/y]_{ys}) \\
  \Updownarrow \mbox{\small (by Prop.~\ref{lem-simp-subst}(3), since 
\coll{\mbox{$(\xs,x) \not= (\ys,y)$ and $\fresh\;\xs\;x\;Y$)}}
} \\
\freshAbs\;\zs\;z\;(\Abs\;\xs\;x\;(X\,[Y/y]_{ys})) \\
  \Updownarrow \mbox{\small (by Prop.~\ref{lem-simp-fresh}(3), since \coll{\mbox{$(\xs,x) \not= (\zs,z)$)}}} \\
\fresh\;\zs\;z\;(X\,[Y/y]_{ys}) \\
  \Updownarrow \mbox{\small (by Induction Hypothesis)} \\
((\zs,z) = (\ys,y) \lor \fresh\;\zs\;z\;X) \,\land\,  (\fresh\;\ys\;y\;X \lor \fresh\;\zs\;z\;Y) \\
  \Updownarrow \mbox{\small (by Prop.~\ref{lem-simp-fresh}(3) applied twice, 
since \coll{\mbox{$(\xs,x) \not= (\zs,z)$}} and \coll{\mbox{$(\xs,x) \not= (\ys,y)$)}} } \\
((\zs,z) = (\ys,y) \lor \freshAbs\;\zs\;z\;(\Abs\;\xs\;x\;X)) \,\land\,  (\freshAbs\;\ys\;y\;(\Abs\;\xs\;x\;X) \lor \fresh\;\zs\;z\;Y)
\end{array} 
$$

The practice of assuming freshness, known in the literature as the Barendregt convention, 
is a hallmark in informal reasoning about bindings. 
Thanks to insight from Nominal Logic 
\cite{pitts-AlphaStructural,UrbanTasson,urban-Barendregt}, 
we also know how to apply this morally correct convention fully rigorously. 
To capture it in our formalization, we 
model parameters 
$p : \param$ as anything that allows for a notion of freshness, or, alternatively, 
provides a set of (free) variables for each varsort, 
$\varsOf : \param \ra \varsort \ra \var\;\sett$. 
With this, a ``fresh induction'' principle can be formulated,  
if all parameters have fewer variables than $|\var|$ (in particular, if they have only finitely many). 

\begin{thm}\label{th-fresh-induct} \rm
Let $\phi : \term \ra \param \ra \bool$ and $\phiAbs : \abs \ra \param \ra \bool$. 
Assume: 
\\(1) $\forall \xs,p.\;|\varsOf\;\xs\;p| < |\var|$
\\(2) $\forall \xs,x,p.\;\phi\;(\Var\;\xs\;x)\;p$
\\(3) $\forall \delta,\inp,\binp,p.\;  
  |\dom\;\inp| < |\var| \,\wedge\, |\dom\;\binp| < |\var|  \,\wedge\,
\lift\,(\lambda X.\;\good\;X \,\wedge\,  (\forall q.\,\phi\;X\;q))\;\allowbreak\inp  \,\wedge\, 
\lift\,(\lambda A.\;\goodAbs\;A \,\wedge\, 
(\forall q.\,\phiAbs\;A\;q))\,\binp \Lra \phi\,(\Op\;\delta\;\inp\;\binp)\;p$
\\(4) $\forall \xs,x,X,p.\;\good\;X \,\wedge\, \phi\;X\;p \,\wedge\, 
\coll{x \not\in \varsOf\;\xs\;p} \,\Lra\, \phiAbs\,(\Abs\;\xs\;x\;X)\;p$

Then $\forall X,p.\;\good\;X \Lra \phi\;X\;p$ and 
$\forall A,p.\;\goodAbs\;A  \Lra \phiAbs\;A\;p$. 
\end{thm} 

Highlighted is the essential difference from the usual structural induction: The bound 
variable $x$ can be assumed fresh for the parameter $p$ (on its varsort, $\xs$). 
Note also that, in the $\Op$ case, we lift to inputs the predicate 
as quantified universally over all parameters.  

\leftOut{ 
To make it easier to instantiate to commonly used parameters, we prove a corollary 
for custom parameters that may contain variables, terms, abstractions and assignments 
of variables to terms:
$$
\textsf{datatype }\param \;=\; \Par\;(\varsort \ra \var\;\llist)\;(\term\;\llist)
\;(\abs\;\llist)\;((\var \ra \term\;\option)\;\llist)
$$
and define $\varsOf\;p$ to take 
}

Back to Prop.~\ref{lem-long}(4), this follows automatically by fresh induction 
(plus the shown simplifications), after recognizing as parameters the variables $(\ys,y)$ and 
$(\zs,z)$ and the term $Y$---formally, 
taking $\param = (\varsort \times \var)^2 \times \term$ and 
$\varsOf\;\xs\;((\ys,y),(\zs,z),\allowbreak Y) = \{y \mid \xs = \ys\} \,\cup\, 
\{z \mid \xs = \zs\}  \,\cup\, 
\{x \mid \neg\,\fresh\;\xs\;x\;Y\}$. 

\subsection{Freshness- and Substitution- Sensitive Recursion}\label{sec-RecDef}

A {\em freshness-substitution (FS) model} consists of two collections of elements endowed 
with term- 
and abstraction- like operators satisfying some 
characteristic properties of terms. 
More precisely, it consists of: 
\begin{myitem}
\item two types, $\T$ and $\A$ 
\item operations corresponding to the 
constructors: $\VAR : \varsort \ra \var \ra \T$, 
$\OP : \opsym \allowbreak\ra (\indexx,\T)\;\inputt \ra (\bindexx,\A)\;\inputt \ra \T$, 
$\ABS : \varsort \ra \var \ra \T \ra \A$
\item operations corresponding to freshness and substitution:
$\FRESH : \varsort \ra \var \ra \T \ra \bool$, $\FRESHABS : \varsort \ra \var \ra \A \ra \bool$, 
$\_[\_/\_]_\_ : \T \ra \T \ra \var \ra \varsort \ra \T$ and 
$\_[\_/\_]_\_ : \A \ra \T \ra \var \ra \varsort \ra \A$
\end{myitem}
and it is required to satisfy the analogues of:
\begin{myitem}
\item the implicational simplification rules for $\fresh$ from Prop.~\ref{lem-imp-fresh} 
\\(for example, $(\ys,y) \not= (\xs,x) \,\Lra\, \FRESH\;\ys\;y\;(\VAR\;\xs\;x)$)
\item the simplification rules for substitution from Prop.~\ref{lem-simp-subst}
\item the substitution-based abstraction equality rules from Prop.~\ref{lem-Abs-subst}
\end{myitem}

\begin{thm}\label{th-rec} \rm
The good terms and abstractions form the initial FS model. Namely, 
for any FS model as above, there exist the functions $\f:\term \ra \T$ 
and $\fAbs : \abs \ra \A$ that commute, on good terms, 
with the constructors and with substitution and preserve freshness: 
$$
\begin{array}{ll}
\f\,(\Var\;\xs\;x) = \VAR\;\xs\;x
&\hspace*{2ex}
\f\,(\Op\;\delta\;\inp\;\binp) = \OP\;\delta\;(\lift\,\f\;\inp)\;(\lift\,\fAbs\;\binp)
\\
\fAbs\,(\Abs\;\xs\;x\;X) = \ABS\;\xs\;x\;(f\;X)
&\hspace*{2ex}
\\
\f\,(X\,[Y/y]_{ys}) = (\f\;X)\,[(\f\;Y)/y]_{ys} 
&\hspace*{2ex}
\fAbs\,(A\,[Y/y]_{ys}) = (\fAbs\;A)\,[(\f\;Y)/y]_{ys} 
\\
\fresh\;\xs\;x\;X \,\Lra\, \FRESH\;\xs\;x\;(\f\;X) 
&\hspace*{2ex}
\freshAbs\;\xs\;x\;A \,\Lra\, \FRESHABS\;\xs\;x\;(\fAbs\;A)
\end{array}
$$
\par
In addition, the two functions are uniquely determined on good terms and abstractions, 
in that, for all other functions $\g:\term \ra \T$ 
and $\gAbs :\abs \ra \A$ satisfying the same commutation and preservation properties, 
it holds that 
$\f$ and $\g$ are equal on good terms and $\fAbs$ and $\gAbs$ are equal on good abstractions.
\end{thm}  

Like any initiality property, this theorem represents a primitive recursion principle. 
Consider first the simpler case of lists over a type $\G$, with constructors $\Nil : \G\;\llist$ 
and $\Cons : \G \ra \G\;\llist \ra \G\;\llist$. To define, by primitive recursion, 
a function from lists, say, $\llength : \G\;\llist \ra \nat$, we need to indicate 
what is $\Nil$ mapped to, here $\llength\;\Nil = 0$, and, recursively, what 
is $\Cons$ mapped to, here $\llength\;(\Cons\;a\;\as) = 1 + \llength\;\as$. 
We can rephrase this by saying: If we define ``list-like'' operators on the target domain---
here, taking $\NIL : \nat$ to be $0$ and $\CONS : \G \ra \nat \ra \nat$ to be $\lambda g,n.\;1+n$---then 
the recursion principle offers us a function $\llength$ that commutes with the constructors: 
$\llength\;\Nil = \NIL = 0$ and $\llength\,(\Cons\;a\;\as) = \CONS\;a\;(\llength\;\as) = 1 + \llength\;\as$. 
For terms, we have a similar situation, except that (1) substitution and freshness 
are considered in addition to the constructors and 
(2) paying the price for lack of freeness, some conditions need to be verified  
to 
deem the operations ``term-like.''  

This recursion principle was discussed in \cite{pop-recPrin} for the 
syntax of $\lambda$-calculus 
and shown to have many useful applications.   
Perhaps the most useful one is the seamless interpretation 
of syntax in semantic domains, in a manner that is guranteed to be 
compatible with alpha, substitution and freshness. We formalize this in our general setting:

A {\em semantic domain} consists of two collections of elements endowed with interpretations 
of the $\Op$ and $\Abs$ constructors, the latter 
in a higher-order fashion---interpreting variable binding as (meta-level) functional binding. 
Namely, it consists of:
\begin{myitem}
\item two types, $\Dt$ and $\Da$ 
\item 
a function $\op : \opsym \ra (\indexx,\Dt)\;\inputt \ra (\bindexx,\Da)\;\inputt \ra \Dt$ 
\item 
 a function $\abss : \varsort \ra (\Dt \ra \Dt) \ra \Da$
\end{myitem}

\begin{thm}\label{th-sem} \rm
The terms and abstractions are 
interpretable in any semantic domain. 
Namely, if $\val$ is the type of valuations of 
variables in the domain, 
$\varsort \ra \var \ra \Dt$, 
there exist the functions 
$\sem : \term \ra \val \ra \Dt$  
and $\semAbs : \abs \ra \val \ra \Da$ such that:
\begin{myitem}
\item $\sem\,(\Var\;\xs\;x)\,\rho = \rho\;\xs\;x$
\item $\sem\,(\Op\;\delta\;\inp\;\binp)\,\rho = \op\;\delta\;
  (\lift\,(\lambda X.\,\sem\;X\;\rho)\,\inp)\;(\lift\,(\lambda A.\,\semAbs\;A\;\rho)\,\binp)$
\item $\semAbs\,(\Abs\;\xs\;x\;X)\,\rho = \abss\;\xs\,(\lambda d.\;\sem\;X\;(\rho[(\xs,x) \la d]))$
\end{myitem}
\par
In addition, the interpretation functions map syntactic substitution and freshness to semantic versions of the 
concepts: 
\begin{myitem}
\item $\sem\,(X[Y/y]_{ys})\,\rho = \sem\;X\;(\rho[(\ys,y) \la \sem\;Y\;\rho])$
\item $\fresh\;\xs\;x\;X \;\Lra\;(\forall \rho,\rho'.\;\rho=_{(xs,x)}\rho' \,\Lra\, \sem\;X\;\rho = \sem\;X\;\rho')$,  
\\where ``$=_{(xs,x)}$'' means equal everywhere but on $(\xs,x)$ 
\end{myitem}
\end{thm}

Theorem~\ref{th-sem} is the foundation for many particular semantic interpretations, including 
that of $\lambda$-terms in Henkin models and that of FOL terms and formulas in FOL models. 
It guarantees compatibility with alpha and proves, as bonuses, a freshness and a substitution 
property.   
The freshness property is nothing but the notion that the interpretation 
only depends on the free variables, whereas the substitution property generalizes 
what is usually called {\em the substitution lemma}, stating that interpreting a substituted term 
is the same as interpreting the original term in a ``substituted'' environment. 

This theorem follows by an instantiation of the recursion Theorem \ref{th-rec}: 
taking $\T$ and $\A$ to be $\val \ra \Dt$ and $\val \ra \Da$ and taking the 
term/abstraction-like operations as prescribed by the desired clauses for $\sem$ and $\semAbs$---e.g., 
$\VAR\;\xs\;x$ is $\lambda \rho.\;\rho\;\xs\;x$. 

\section{Sorting the Terms}  \label{sec-sorting}

So far, we have 
a framework where the operations take as 
free and bound inputs partial families of terms and abstractions.  
All theorems refer to good (i.e., sufficiently 
low-branching) terms and abstractions. 
However, we promised a theory that is applicable to terms over many-sorted binding signatures. 
Thanks to the choice of a flexible notion of input, it is not hard 
to cast our results into such a many-sorted setting. 
Given a suitable 
notion of signature (\ref{subsec-sign}), 
we classify terms according to sorts (\ref{subsec-termsSig}) 
and prove that well-sorted terms are good (\ref{subsec-goodToSort})---this 
gives us sorted versions of all theorems (\ref{subsec-endProd}).

\subsection{Binding Signatures} \label{subsec-sign}

%

A {\em (binding) signature} is a tuple
$(\indexx,\bindexx,\varsort,\sort,\opsym,\asSort,\stOf,\arOf,\allowbreak\barOf)$, 
where $\indexx$, $\bindexx$, $\varsort$ and $\opsym$ are types (with the previously discussed intuitions) and 
$\sort$ is a new type, of sorts for terms. Moreover: 
\begin{myitem}
\item $\asSort : \varsort \ra \sort$ is an injective map, embedding varsorts into sorts
\item $\stOf : \opsym \ra \sort$, read ``the (result) sort of''
\item $\arOf : \opsym \ra (\indexx,\sort)\,\inputt$, read ``the (free) arity of"
\item $\barOf : \opsym \ra (\bindexx,\varsort \times \sort)\,\inputt$, read ``the bound arity of" 
\end{myitem}

Thus, 
a signature prescribes which varsorts correspond to which sorts (as discussed in Section~\ref{prel-manySorted}) 
and, for each operation symbol, 
which are the sorts of its free inputs (the arity), of its bound (abstraction) inputs (the bound arity),  
and of its result.

When we give examples for our concrete syntaxes in Section~\ref{sec-exa}, 
we will write $(i_1\mapsto a_1,\ldots,i_n\mapsto a_n)$ for the partial 
function that sends each $i_k$ to $\Some\;a_k$ and everything else to $\None$. In particular, $()$ 
denotes the totally undefined function. 

For the $\lambda$-calculus syntax, we take $\indexx = \bindexx = \nat$, $\varsort = \sort = \{\lamterm\}$ (a singleton datatype), 
$\opsym = \{\App,\Lam\}$, 
$\asSort$ to be the identity and $\stOf$ to be the unique function to $\{\lamterm\}$. 
Since $\App$ has two free inputs and no bound input, we use the first two elements of $\nat$ as free arity 
and nothing for the 
bound arity:
$\arOf\;\App =(0\mapsto \lamterm,\,1\mapsto\lamterm)$, 
$\barOf\;\App = ()$. 
By contrast, since $\Lam$ has no free input and one bound input, we use nothing for the 
free arity, and the first element of $\nat$ for the bound arity: 
$\arOf\;\Lam = ()$, 
$\barOf\;\Lam = (0\mapsto (\lamterm,\lamterm))$. 

For the CCS example in Section~\ref{prel-infBranch}, 
we fix a type $\chan$ of channels. 
We choose a cardinal upper bound $\kappa$ for the branching of sum ($\sum$), and choose a type 
$\indexx$ of cardinality $\kappa$. 
For $\bindexx$, we do not need anything special, so we take it to be $\nat$. 
We have two sorts, of expressions and processes, 
so we take $\sort = \{\exp,\proc\}$.  
Since we have expression variables but no process variables, we take $\varsort = \{\varexp\}$ and $\asSort$ to send 
$\varexp$ to $\exp$. 
We define $\opsym$ as the following datatype:
%
$\opsym \; =\; \Zero \mid \Plus \mid \Inp\;\chan \mid \Out\;\chan \mid {\textstyle \sum}\,(\indexx\;\sett)$. 
%
%
The free and bound arities and sorts of the operation symbols are as expected. For example, 
$\Inp\;c$ acts 
similarly to $\lambda$-abstraction, but binds, in $\proc$ terms, variables of a different sort, $\varexp$: 
$\arOf\,(\Inp\;c) = ()$, 
$\barOf\,(\Inp\;c) = (0\mapsto (\varexp,\proc))$. 
For $\sum I$ with $I:\indexx\;\sett$, the arity is only defined for elements of $I$, namely  
$\arOf\;(\sum I) = ((i\in I)\mapsto \proc)$. 

\subsection{Well-Sorted Terms Over a Signature}
\label{subsec-termsSig}

Based on the information from a signature, we can distinguish our terms of interest, 
namely those that are well-sorted in the sense that:
\begin{myitem}
\item all 
variables are embedded into terms of sorts compatible with their varsorts 
\item all 
operation symbols are applied according their free and bound arities
\end{myitem}

This is modeled by well-sortedness predicates $\wls: \sort \ra \term \ra \bool$ and 
$\wlsAbs: \varsort \ra \sort \ra \abs \ra \bool$, where $\wls\;s\;X$ 
states that $X$ is a well-sorted term of sort $s$ and $\wlsAbs\;(\xs,s)\;A$ 
states that $A$ is a well-sorted abstraction binding an $\xs$-variable in an $s$-term. 
They are defined mutually inductively by the 
following clauses: 
$$\begin{array}{rcl}
&&\wls\;(\asSort\;\xs)\;(\Var\;\xs\;x)
\\
\lift\,\wls\;(\arOf\;\delta)\;\inp \,\wedge\, \lift\,\wlsAbs\;(\barOf\;\delta)\;\binp &\;\LRA\;&
 \wls\;(\stOf\;\delta)\;(\Op\;\delta\;\inp\;\binp)
\\
\isInBar\,(\xs,s) \,\wedge\, \wls\;s\;X &\;\LRA\;& \wlsAbs\;(\xs,s)\;(\Abs\;\xs\;x\;X)
\end{array}
$$
where $\isInBar\,(\xs,s)$ states that the pair $(\xs,s)$ is in the bound arity of at 
least one operation symbol $\delta$, i.e., $\barOf\;\delta\;i = (\xs,s)$ for some $i$---
this 
rules out unneeded abstractions. 


%
Let us illustrate sorting for our running examples. 
In the $\lambda$-calculus syntax, 
let $X = \Var\;\lam\;x$, 
$A = \Abs\;\lam\;x\;X$, and 
$Y = \Op\;\Lam\;()\;(0 \mapsto A)$. 
These correspond to what, in the unsorted BNF notation from Section~\ref{prel-abs}, 
we would write $\Var\;x$, $\Abs\;x\;X$ and $\Lam\;(\Abs\;x\;X)$. 
In our sorting system, $X$ and $Y$ are both well-sorted terms at sort $\lam$ 
(written $\wls\;\lam\;X$ and $\wls\;\lam\;Y$) and 
$A$ is a well-sorted abstraction at sort $(\lam,\lam)$ (written $\wlsAbs\,(\lam,\lam)\,A$). 

For CCS, we have that 
$E = \Op\;\Zero\,()\,()$ and 
$F = \Op\;\Plus\;(0\mapsto E,\;1\mapsto E)\,()$ are well-sorted terms of sort $\exp$. 
Moreover, $P = \Op\,(\sum \emptyset)\,()\,()$ and 
$Q = \Op\,(\Out\;c)\,\allowbreak(0\mapsto F,1 \mapsto P)\,()$ are well-sorted terms of sort $\proc$. 
(Note that $P$ is a sum over the empty set of choices, i.e., the null process, whereas 
$Q$ represents a process that outputs the value of $0+0$ on channel $c$ and then stops.)
If, e.g., 
we swap the arguments of $\Out\;c$ in $Q$, we obtain 
$\Op\,(\Out\;c)\,(0 \mapsto P,1\mapsto F)\,()$, which is not well-sorted: 
In the inductive clause for $\wls$, the input $(0 \mapsto P,1\mapsto F)$ fails 
to match the arity of $\Out\;c$, 
$(0 \mapsto \exp,1\mapsto \proc)$.

\subsection{From Good to Well-Sorted}
\label{subsec-goodToSort}
  
Recall that goodness means ``does not branch beyond $|\var|$.'' On the other hand, 
well-sortedness imposes that, for each applied operation symbol $\delta$, 
its inputs have same domains, 
i.e., {\em  only branch as much}, as the arities of $\delta$. Thus, it suffices 
to assume the arity domains smaller than $|\var|$. 
We will more strongly assume that the types of sorts and indexes 
(the latter subsuming the arity domains) 
are 
all smaller than $|\var|$: 

\begin{ass}\rm \label{ass-varLarge}
$|\sort| < |\var| \;\wedge\; |\indexx| < |\var| \;\wedge\; |\bindexx| < |\var|$ 
\end{ass}

Now we can prove:
\begin{prop} \label{lem-wls-good}
$(\wls\;s\;X \Lra \good\;X) \;\wedge\; (\wls\;(\xs,s)\;A \Lra \goodAbs\;A)$
\end{prop}

In addition, we prove that all the standard operators preserve well-sortedness. 
For example, we prove that if we substitute, in the well-sorted term $X$ of sort $s$, 
for the variable $y$ of varsort $\ys$, the well-sorted term $Y$ of sort 
corresponding to $\ys$, then we obtain a well-sorted term of sort $s$: 
$\wls\;s\;X \;\wedge\; \wls\;(\asSort\;\ys)\;Y \;\Lra\; \wls\;s\;(X\,[Y/y]_{ys})$. 

Using the preservation properties and 
Prop.~\ref{lem-wls-good}, we transfer the entire  
theory 
of Sections \ref{subsec-termsTh} and \ref{sec-reas} from good terms to well-sorted terms---e.g., 
Prop.~\ref{lem-long}(2) becomes: 
$$
\begin{array}{c}
\coll{\wls\;s\;X \,\wedge\, \wls\;(\asSort\;\ys)\;Y_1 \,\wedge\, \wls\;(\asSort\;\ys)\;Y_2} 
\Lra
X\,[Y_1 / y]_{ys}\,[Y_2 / y]_{ys} = \ldots 
\end{array}
$$

The transfer is mostly straightforward for all facts, including the induction theorem. 
(For stating the well-sorted version of the recursion and semantic interpretation theorems, 
there is some additional bureaucracy since we also need sorting predicates on the target domain
---Appendix~\ref{app-RecDef} gives details.)

There is an important remaining question: Are our two Assumptions 
(\ref{ass-reg} and \ref{ass-varLarge}) satisfiable? That is, 
can we find, for any types $\sort$, $\indexx$ and $\bindexx$, 
a type $\var$ larger than these such that $|\var|$ is regular? 
Fortunately, the theory of cardinals 
again provides us with a positive answer: 
Let $\G = \nat + \sort + \indexx + \bindexx$. 
Since 
any successor of an infinite cardinal 
is regular, we can 
take $\var$ to have the same cardinality as the successor of $|\G|$, by defining 
$\var$ as a suitable subtype of $\G\;\sett$. 
In the case of all operation symbols being finitary, i.e., with 
their arities having finite domains, we do not need the above fancy construction,  
but can simply take $\var$ to be a copy of $\nat$.  


\subsection{End Product}
\label{subsec-endProd}

All in all, 
our formalization provides a theory of syntax with bindings over an arbitrary many-sorted signature.
The signature is formalized as an Isabelle locale \cite{Locales} 
that fixes 
the types $\var$, $\sort$, $\varsort$, $\indexx$, $\bindexx$ and $\opsym$ 
and the constants $\asSort$, $\arOf$ and $\barOf$ and 
assumes the injectivity of $\asSort$ and the $\var$ 
properties (Assumptions \ref{ass-reg} and \ref{ass-varLarge}).     
All end-product theorems are placed 
in this locale.

The whole formalization 
consists of 22700 lines of code (LOC). Of these, 3300 LOC are dedicated 
to quasiterms, their standard operators and alpha-equivalence. 
3700 LOC are dedicated to the definition of terms and the lifting of results from quasiterms.
Of the latter, the properties of substitution were the most extensive---2500 LOC out of 
the whole 3700---since substitution, unlike freshness and swapping, 
requires heavy variable renaming, which complicates the proofs. 

The induction and recursion schemes presented in Section~\ref{sec-reas} are  
not the only schemes we formalized (but are the most useful ones). 
We also proved a variety of lower-level induction schemes 
based on the skeleton of the terms (a generalization of depth for possibly infinitely branching terms) 
and schemes that are easier to instantiate---e.g., by pre-instantiating Theorem~\ref{th-fresh-induct}   
with commonly used parameters such as variables, terms and environments. 
As for the recursion Theorem~\ref{th-rec}, we additionally proved a more flexible scheme that allows 
the recursive argument, and not only the recursive result, to be referred---this is 
{\em full-fledged primitive recursion}, whereas Theorem \ref{th-rec} only implements {\em iteration}. 
Also, we proved schemes for recursion that factor swapping \cite{norrish-MechanisingLambdaInFirstOrder} 
instead of and in addition to substitution. 
All together, these constitute 8000 LOC. 

The remaining 7700 LOC of the formalization are dedicated to transiting from good terms 
to sorted terms. Of these, 3500 LOC are taken by the sheer statement 
of our many end-product theorems. Another fairly large part, 2000 LOC, is dedicated to 
transferring all the variants of the recursion 
Theorem~\ref{th-rec} and the interpretation Theorem~\ref{th-sem}, which require conceptually straightforward
but technically tedious moves back and forth 
between sorted terms and sorted elements of the target domain.   


\section{Discussion,  
Related Work and Future Work} \label{sec-RelWork}

There is a large amount of literature on formal approaches to syntax with bindings. 
(See \cite[\S2]{POPLmark}, \cite[\S6]{momFelty-Hybrid4}  and \cite[\S2.10,\S3.7]{pop-thesis} for overviews.)  
%
Our work, nevertheless, fills a gap in the literature: 
It is the first theory of binding syntax mechanized in 
a universal algebra fashion, i.e., with sorts and many-sorted term constructors specified 
by a binding signature,  
as employed in several theoretical developments, e.g., \cite{fio-abs,pitts-AlphaStructural,sun-alg,DBLP:journals/tcs/0001R15}.  
The universal algebra aspects 
of our approach are the consideration of an {\em arbitrary signature} and the singling out of 
the collection of terms and the operations on them 
as an {\em initial object in a category of models/algebras} (which yields a recursion principle). 
We do not consider arbitrary equational varieties (like in \cite{sun-alg}), 
but only focus on selected equations and Horn clauses that characterize the term 
models (like in \cite{pitts-AlphaStructural}). 

{\bf Alternatives to Universal Algebra}  
A popular alternative to our universal algebra approach is higher-order abstract syntax (HOAS) 
\cite{har-fra,weakHOAS,momFelty-Hybrid4,chlipala-Parametric,gun-proper}: 
the reduction of all 
bindings to a single binding---that of a fixed $\lambda$-calculus. 
Compared to universal algebra, HOAS's 
advantage is lighter formalizations, 
whereas the disadvantage is the need to prove the representation's adequacy (which 
involves reasoning about substitution) and, in some frameworks, 
the need to rule out the resulted junk (also known as exotic terms).  

Another alternative, very successfully used in HOL-based provers such as HOL4 \cite{slind-norrish-2008} and Isabelle/HOL, 
is the ``package'' approach: Instead of deeply embedding sorts and 
operation symbols like we do, packages take a user a specification 
of the desired types and operations and prove all the theorems for that instance (on a dynamic basis).  
Nominal Isabelle \cite{urban-2008,urbanGeneralBinders} is 
a popular such package, which implements terms with bindings for Isabelle/HOL. 
From a theoretical perspective, a universal algebra theory has a wider appeal, as it models 
``statically'' the meta-theory in its whole generality. However, a package 
is more practical, since most proof assistant users only care about the particular instance 
syntax used in their development. 
In this respect, simply instantiating our signature with the particular syntax is 
not entirely satisfactory, since it is not sufficiently ``shallow''---e.g., one would like to have actual operations such as $\Lam$     
instead of applications of $\Op$ to a $\Lam$ operation symbol, 
and would like to have actual types, such as $\expp$ and $\procc$, instead 
of the well-sortedness predicate applied to sorts, $\wls\;\exp$ and $\wls\;\proc$. 
For our applications, so far we have 
manually transited from our ``deep'' signature instances 
to the more usable shallow version sketched above.  
In the future, we plan to have this transit process automated, obtaining the best of both worlds, 
namely a universal algebra theory that also acts as a {\em statically certified} package. 
(This approach has been prototyped for a 
smaller theory: that of 
nonfree equational datatypes \cite{schropp-nonfree}.)

{\bf Theory of Substitution and Semantic Interpretation}  
%
The main goal of our work was the development of 
as much as possible from the theory of syntax 
for an arbitrary syntax. 
To our knowledge, 
none of the existing frameworks provides 
support for substitution 
and the interpretation of terms in semantic domains at this level of generality. 
Consequently, formalizations for concrete syntaxes, even those based on 
sophisticated packages such as Nominal Isabelle or the similar 
tools and formalizations in Coq \cite{nominalCoq,aydemirPOPL08,Hirschowitz:2012}, have to redefine these standard concepts 
and prove their properties over and over again---an unnecessary consumption 
of time and brain power.

{\bf Induction and Recursion Principles}   
There is a rich literature on these topics, which are connected to the quest, 
pioneered by 
Gordon and Melham \cite{gor-5axAlpha}, 
of understanding terms with bindings modulo alpha as an abstract datatype. 
We formalized the Nominal structural induction principle 
from \cite{pitts-AlphaStructural}, which is also implemented in Nominal Isabelle. 
By contrast, we did not go after 
the Nominal recursion principle. Instead, we chose to stay more faithful to the 
abstract datatype desideratum,  
generalizing to an arbitrary syntax 
our own schema for substitution-aware recursion \cite{pop-recPrin} and  
Michael Norrish's schema for swapping-aware recursion \cite{norrish-MechanisingLambdaInFirstOrder}---both 
of which can be read as stating that terms with bindings are Horn-abstract datatypes, i.e., 
are initial models of certain Horn theories 
\cite[\S3,\S8]{pop-recPrin}. 

{\bf Generality of the Framework} 
Our constructors are restricted to binding at most one variable 
in each input---a limitation that makes our framework far from ideal for representing 
complex binders such as the let patterns of POPLmark's Challenge 2B.     
In contrast, the specification language Ott \cite{ott-tool} 
and 
Isabelle's Nominal2 package \cite{urbanGeneralBinders} were specifically designed 
to address such complex, possibly recursive binders.  
Incidentally, the Nominal2 package also separates abstractions from terms, like we do, but their abstractions 
are significantly more expressive; their terms are also quotiented to  
alpha-equivalence, which is defined via flattening the binders into 
finite sets or lists of variables (atoms).   
      
On the other hand, to the best of our knowledge, our formalization is the first to capture infinitely branching terms 
and our foundation of alpha equivalence on the regularity of $|\var|$ 
is also a theoretical novelty---constituting a less exotic alternative 
to Murdoch Gabbay's work on infinitely supported objects in nonstandard set theory \cite{Gabbay:2007}.
This flexibility would be needed to formalize 
calculi such as  
infinite-choice process algebra, for which 
infinitary structures 
have been previously employed to give semantics \cite{Lut02}.

{\bf Future Generalizations and Integrations}  
Our theory currently addresses mostly {\em structural} aspects of terms. A next step would be 
to cover {\em behavioral} aspects, such as formats for SOS rules and their interplay with binders, 
perhaps building on existing Isabelle formalizations of process algebras and programming languages 
(e.g., \cite{pop-coind,prob-nonint,conc-nonint,DBLP:conf/popl/NipkowO98,lochbihler-2012,psiInIsa}). 
  
Another exciting prospect is the integration of our framework with Isabelle's recent  
package for inductive and coinductive datatypes \cite{blanchette-et-al-2014-tru} 
based on bounded natural functors (BNFs),  
which follows a compositional design \cite{traytel-et-al-2012}
and provides flexible ways to nest types \cite{nonuniform-lics2017} and mix recursion with 
corecursion \cite{fouco,amico}, but does not yet cover terms with bindings. 
Achieving compositionality in the presence of bindings will require a substantial 
refinement of the notion of BNF (since terms with bindings form only partial functors w.r.t.\ their sets of free variables).

\leftOut{
\ \\
{\bf Conclusion} 
We presented 
the first mechanization  
of a theory of bindings over a many-sorted signature. 
This formed the basis of several applications of 
formalized meta-theory for concrete syntaxes,    
which 
\href{http://curiosamathematica.tumblr.com/post/151061725677/q-what-do-you-call-someone-reading-a-category}
{so far have been developed by the second author 
and his coauthors. 
We hope that the presentation of our framework 
and its underlying principles and design decisions 
will facilitate its use by non-coauthors too.}
}

\smallskip
\paragraph*{\small Acknowledgment}  

\small
We thank the
anonymous reviewers for suggesting textual improvements.
Popescu has received funding from UK's Engineering and Physical Sciences Research Council 
(EPSRC) via the grant EP/N019547/1, Verification of Web-based  Systems (VOWS).

\bibliographystyle{splncs03}
\bibliography{bib}{}


\include{Appendix}

\end{document}

%% file: myCommands.tex
\newtheorem{thm}[lemma]{Theorem}
\newtheorem{prop}[lemma]{Prop}
\newtheorem{conv}[lemma]{Convention}
\newtheorem{ass}[lemma]{Assumption}

\newcommand\leftOut[1]{}

\newcommand{\ov}{\overline}

\renewcommand{\phi}{\varphi}

\renewcommand{\iff}{\Longleftrightarrow}

\renewcommand{\partial}{\rightharpoondown}
\newcommand{\la}{\leftarrow}

\newcommand{\ra}{\rightarrow}

\newcommand{\Ra}{\Rightarrow}

\newcommand{\Lra}{\Longrightarrow}
\newcommand{\LRA}{\Longrightarrow}

\newcommand{\lamterm}{{\mbox{\rm\textsf{\small lam}}}}
\newcommand{\lam}{{\mbox{\rm\textsf{\small lam}}}}

\newcommand\abstraction{quasiabstraction}
\newcommand\abstractions{quasiabstractions}

\newcommand{\abss}{\mbox{\rm\textsf{\small abs}}}
\newcommand{\Suc}{\mbox{\rm\textsf{\small Suc}}}
\newcommand{\pred}{\mbox{\rm\textsf{\small pred}}}
\newcommand{\Branch}{\mbox{\rm\textsf{\small Branch}}}
\newcommand{\qSkel}{\mbox{\rm\textsf{\small qSkel}}}
\newcommand{\qSkelAbs}{\mbox{\rm\textsf{\small qSkelAbs}}}
\newcommand{\skel}{\mbox{\rm\textsf{\small skel}}}
\newcommand{\skelAbs}{\mbox{\rm\textsf{\small skelAbs}}}
\newcommand{\sem}{\mbox{\rm\textsf{\small sem}}}
\newcommand{\semAbs}{\mbox{\rm\textsf{\small semAbs}}}
\newcommand{\proc}{\mbox{\rm\textsf{\small proc}}}
\newcommand{\varexp}{\mbox{\rm\textsf{\small varexp}}}
\renewcommand{\exp}{\mbox{\rm\textsf{\small exp}}}

\newcommand{\proj}{\mbox{\rm\textsf{\small proj}}}
\newcommand{\projAbs}{\mbox{\rm\textsf{\small projAbs}}}
\newcommand{\good}{\mbox{\rm\textsf{\small good}}}
\newcommand{\qGood}{\mbox{\rm\textsf{\small qGood}}}
\newcommand{\dom}{\mbox{\rm\textsf{\small dom}}}
\newcommand{\goodAbs}{\mbox{\rm\textsf{\small goodAbs}}}
\newcommand{\qGoodAbs}{\mbox{\rm\textsf{\small qGoodAbs}}}
\newcommand{\al}{\mbox{\rm\textsf{\small alpha}}}
\newcommand{\alAbs}{\mbox{\rm\textsf{\small alphaAbs}}}

\newcommand{\lift}{\mbox{$\uparrow$}}

\newcommand{\ABS}{\mbox{\rm\textsf{\small ABS}}}

\newcommand{\varsOf}{\mbox{\rm \textsf{\small vars\hspace{-0.1ex}Of}}}
\newcommand{\None}{\mbox{\rm\textsf{\small None}}}
\newcommand{\Some}{\mbox{\rm\textsf{\small Some}}}
\newcommand{\wls}{\mbox{\rm\textsf{\small wls}}}
\newcommand{\wlsAbs}{\mbox{\rm\textsf{\small wlsAbs}}}

\newcommand{\Inp}{\mbox{\rm\textsf{\small Inp}}}
\newcommand{\Out}{\mbox{\rm\textsf{\small Out}}}

\newcommand{\isInBar}{\mbox{\rm\textsf{\small isInBar}}}

\newcommand{\qOp}{\mbox{\rm\textsf{\small q\hspace*{-0.1ex}Op}}}

\newcommand{\qAbs}{\mbox{\rm\textsf{\small q\hspace*{-0.1ex}Abs}}}
\newcommand{\asSort}{\mbox{\rm\textsf{\small asSort}}}

\newcommand{\stOf}{\mbox{\rm\textsf{\small stOf}}}

\newcommand{\arOf}{\mbox{\rm\textsf{\small arOf}}}

\newcommand{\barOf}{\mbox{\rm\textsf{\small barOf}}}

\newcommand{\Ttrue}{\mbox{\rm\textsf{\small True}}}
\newcommand{\Ffalse}{\mbox{\rm\textsf{\small False}}}
\newcommand{\Var}{\mbox{\rm\textsf{\small Var}}}
\newcommand{\qVar}{\mbox{\rm\textsf{q\hspace*{-0.2ex}Var}}}

\newcommand{\App}{\mbox{\rm\textsf{\small App}}}

\newcommand{\Lam}{\mbox{\rm\textsf{\small Lam}}}

\newcommand{\Lm}{\mbox{\rm\textsf{\small Lm}}}

\newcommand{\Abs}{\mbox{\rm\textsf{\small Abs}}}

\newcommand{\fresh}{\mbox{\rm\textsf{\small fresh}}}
\newcommand{\FRESH}{\mbox{\rm\textsf{\small FRESH}}}

\newcommand{\qFresh}{\mbox{\rm\textsf{\small q\hspace*{-0.1ex}Fresh}}}
\newcommand{\qFreshAbs}{\mbox{\rm\textsf{\small qFreshAbs}}}

\newcommand{\freshAbs}{\mbox{\rm\textsf{\small freshAbs}}}
\newcommand{\FRESHABS}{\mbox{\rm\textsf{\small FRESHABS}}}

\newcommand{\Par}{\mbox{\rm\textsf{\small Par}}}

\newcommand{\op}{\mbox{\rm\textsf{\small op}}}
\newcommand{\Op}{\mbox{\rm\textsf{\small Op}}}
\newcommand{\OP}{\mbox{\rm\textsf{\small OP}}}

\newcommand{\llength}{\mbox{\rm\textsf{\small length}}}
\newcommand{\Cons}{\mbox{\rm\textsf{\small Cons}}}
\newcommand{\Nil}{\mbox{\rm\textsf{\small Nil}}}
\newcommand{\CONS}{\mbox{\rm\textsf{\small CONS}}}
\newcommand{\NIL}{\mbox{\rm\textsf{\small NIL}}}

\newcommand{\Zero}{\mbox{\rm\textsf{\small Zero}}}
\newcommand{\Plus}{\mbox{\rm\textsf{\small Plus}}}

\newcommand{\VAR}{\mbox{{\rm \textsf{\small V$\hspace*{-0.1ex}$A$\hspace*{-0.1ex}$R}}}}

\newcommand{\phiAbs}{\mbox{$\phi$\textit{Abs}}}
\newcommand{\f}{\mbox{\textit{f}}}
\newcommand{\fAbs}{\mbox{\textit{fAbs}}}
\newcommand{\g}{\mbox{\textit{g}}}
\newcommand{\gAbs}{\mbox{\textit{gAbs}}}

\newcommand{\inp}{\mbox{\textit{inp}}}
\newcommand{\binp}{\mbox{\textit{binp}}}

\newcommand{\as}{\mbox{\textit{as}}}

\newcommand{\xs}{\mbox{\textit{xs}}}
\newcommand{\ys}{\mbox{\textit{ys}}}
\newcommand{\zs}{\mbox{\textit{zs}}}

\newcommand{\val}{\bf val}
\newcommand{\tre}{\bf tree}
\newcommand{\A}{\bf A}
\newcommand{\T}{\bf T}

\newcommand{\G}{\bf G}
\newcommand{\Dt}{\bf Dt}
\newcommand{\Da}{\bf Da}
\newcommand{\chan}{\mbox{\bf chan}}
\newcommand{\nat}{\mbox{\bf nat}}

\newcommand{\expp}{\mbox{\bf exp}}
\newcommand{\procc}{\mbox{\bf proc}}

\newcommand{\varsort}{\mbox{\bf varsort}}
\newcommand{\sort}{\mbox{\bf sort}}

\newcommand{\bool}{\mbox{\bf bool}}

\newcommand{\var}{\mbox{\bf var}}

\newcommand{\term}{\mbox{\bf term}}

\newcommand{\abs}{\mbox{\bf abs}}

\newcommand{\qterm}{\mbox{\bf qterm}}

\newcommand{\qabs}{\mbox{\bf qabs}}

\newcommand{\indexx}{\mbox{\bf index}}

\newcommand{\bindexx}{\mbox{\bf bindex}}

\newcommand{\param}{\mbox{\bf param}}
\newcommand{\opsym}{\mbox{\bf opsym}}

\newcommand{\option}{\mbox{\bf option}}

\newcommand{\inputt}{\mbox{\bf input}}

\newcommand{\llist}{\mbox{\bf list}}
\newcommand\thmcontinues[1]{Continued}

\newcommand{\sett}{\mbox{\bf set}}

%% file: Appendix.tex
\appendix

\noindent
{\Large APPENDIX}

\ \\ 
This appendix is included for the reviewers' convenience. 
If the paper is accepted, we will replace the references to the appendix with 
references to a technical report made available online.


\section{More on Fresh Induction} \label{app-moreInd}

Fresh induction is based on the possibility to 
rename bound variables in abstractions without loss of generality. 
Intuitively, this is correct since the ``skeleton'' of a term, i.e., the tree 
obtained from it by retaining only branching information and forgetting 
about the operation symbols and the leaf variables, remains the same. 

Namely, we define trees to be ``pure branching'' structures: 
$$
\textsf{datatype }(\alpha,\beta)\tre \;=\; 
\Branch\;((\alpha,\,(\alpha,\beta)\,\tre)\,\inputt)\;((\beta,\,(\alpha,\beta)\,\tre)\,\inputt)
$$ 

The skeleton functions,  
$\qSkel : \qterm \ra (\indexx,\bindexx)\,\tre$ and 
$\qSkelAbs : \qabs \ra (\indexx,\bindexx)\,\tre$, 
are first defined for quasiterms and quasiabstractions by usual structural recursion: 
$$
\begin{array}{rcl}
\qSkel\;(\qVar\;\xs\;x) &\;=\;&
   \Branch\;(\lambda i.\,\None)\;(\lambda i.\,\None)
\\
\qSkel\;(\qOp\;\delta\;\inp\;\binp) &\;=\;&
   \Branch\;(\lift\,\qSkel\;\inp)\;(\lift\,\qSkel\;\binp)
\\
\qSkelAbs\;(\qAbs\;\xs\;x\;X) &\;=\;&
    \Branch\;(\lambda i.\;\Some\,(\qSkel\;X))\;(\lambda i.\,\None)
\end{array}
$$

We show that the skeleton is invariant under alpha-equivalence, which allows 
us to lift these functions to terms and abstractions, 
$\skel : \term \ra (\indexx,\bindexx)\,\tre$ and 
$\skelAbs : \abs \ra (\indexx,\bindexx)\,\tre$. 

Thus, the skeleton gives a measure for terms that generalizes 
the standard depth for finitely branching datatypes. 
Importantly, the skeleton is also invariant under swapping and variable-for-variables substitution. 
This allows us to prove our fresh induction schema 
by standard structural induction on trees, using $\skel$ and $\skelAbs$ as measures.

\section{More on Recursion} \label{app-moreInd}

\subsection{Many-Sorted Recursion}\label{app-RecDef}

As mentioned in the main paper, adapting the theorems from 
good items to well-sorted items is a routine process. 
For recursion, this is more bureaucratic, since it involves 
the sorting of the target domain as well.

A {\em sorted FS model} is an extension 
of the 
FS models from the main paper with
\begin{myitem}
\item the sorting predicates $\wls^{\T} : \sort \ra \T \ra \bool$ and 
$\wlsAbs^{\T} : \varsort \ra \sort \ra \A \ra \bool$
\item the assumption that all operators preserve sorting, 
e.g., 
\\$\wls^{\T}\,s\;X \wedge \wls^{\T}\;(\asSort\;\ys)\;Y 
\;\Lra\;
\wls^{\T}\,s\;X\ [Y / y]_{ys}$
\item the assumption that the sets of items corresponding to different 
sorts are disjoint
\end{myitem}

The recursion theorem is modified to take sorting into account: 

\begin{thm}\label{th-rec-sort} \rm
The sorted terms and abstractions form the initial sorted FS model. Namely, 
for any sorted FS model. 
there exist the functions $\f:\term \ra \T$ 
and $\fAbs : \abs \ra \A$ that satisfy the same properties as in Theorem~\ref{th-rec} 
and additionally preserve sorting:
\begin{myitem}
\item $\wls\;s\;X \;\Lra\; \wls^{\T}\,s\;(\f\;X)$
\item $\wls\;s\;X \;\Lra\; \wlsAbs^{\T}\,s\;(\fAbs\;X)$
\end{myitem}
\end{thm}

Similarly, for a sorted version of the semantic interpretation theorem, we define 
a {\em sorted semantic domain} to have the same components as a semantic domain 
from the main paper, plus sorting predicates $\wls^{\Dt}$ and $\wls^{\Da}$.  
Again, it is assumed that the semantic operators preserve sorting and the 
sets of items for different sorts are disjoint.
Then Theorem~\ref{th-sem} is adapted to sorted domains, additionally ensuring 
the sort preservation of $\sem$ and $\semAbs$.

\subsection{Swapping-Based Recursion}

The main feature of our recursion theorem is the ability to define 
functions in a manner that is compatible with alpha-equivalence. 
A byproduct of the theorem is that the defined functions  
interacts well with freshness and substitution, in that 
it maps these concepts to corresponding concepts on the target domain.  
Michael Norrish has developed a similar principle that employs swapping instead 
of substitution \cite{norrish-MechanisingLambdaInFirstOrder}. We have formalized this 
as well:

A {\em fresh-swapping model} is a structure similar to our fresh-substitution models, 
just that instead of the substitution-like operators, $\_[\_/\_]_\_ : \T \ra \T \ra \var \ra \varsort \ra \T$ and 
$\_[\_/\_]_\_^{Abs} : \A \ra \T \ra \var \ra \varsort \ra \A$, 
it features swapping-like operators, 
$\_[\_\wedge\_]_\_ : \T \ra \var \ra \var \ra \varsort \ra \T$ and 
$\_[\_\wedge \_]_\_^{Abs} : \A \ra \var \ra \var \ra \varsort \ra \A$, assumed to 
satisfy rules corresponding to those for simplifying term swapping: 
\begin{myitem}
\item $(\VAR\;\xs\;x)\,[z_1 \wedge z_2]_{zs} \;=\; \VAR\;\xs\,(x\,[z_1 \wedge z_2]_{xs,zs})$
\item $(\OP\;\delta\;\inp\;\binp)\,[z_1 \wedge z_2]_{zs} = 
\Op\;\delta\;(\lift\,(\_[z_1 \wedge z_2]_{zs})\,\inp)\;(\lift\,(\_[z_1 \wedge z_2]_{zs})\,\binp)$ 
%
%
\item $(\Abs\;\xs\;x\;X)\,[z_1 \wedge z_2]_{zs} \,=\, \Abs\;\xs\,(x\,[z_1 \wedge z_2]_{xs,zs})$
\end{myitem}
and, instead of the substitution-based variable-renaming property, the following 
swapping-based alternative: $\FRESH\;\xs\;y\;X \;\Lra\; \Abs\;\xs\;x\;X = \Abs\;\xs\;y\;(X\,[y \wedge x]_{ys})$. 
Then a version of the recursion theorem holds, in that terms form the initial 
freshness-swapping model. 

In fact, we can combine both notions, obtaining 
freshness-substitution-swapping models, among which again terms form the initial 
model. Having formalized all these variants, the user can decide on the desired 
``contract:'' with more operators factored in, there are more proof obligations that 
need to be discharges for the definition to succeed, but then the 
defined functions satisfy more desirable properties. 

\subsection{Recursion versus Iteration}

What we presented so far was actually a simple version of primitive recursion 
called {\em iteration}. The difference between the two is illustrated by 
the following simple example:\footnote{This is a contrived example, where no ``real'' recursion 
occurs---but 
it illustrates the point.} 
The predecessor function $\pred : \nat \ra \nat$ 
is defined by $\pred\;0 = 0$ and $\pred\,(\Suc\;n) = n$.  
This does not fit the iteration scheme, where only the value of the function on 
smaller arguments, and not the arguments themselves, can be used. 
In the example, iteration would allow 
$\pred\;(\Suc\;n)$ to invoke recursively $\pred\;n$, but not $n$.  
Of course, we can simulate recursion by iteration if we are allowed an auxiliary 
output: defining $\pred' : \nat  \ra \nat \times \nat$ by iteration, 
$\pred'\;0 = (0,0)$ and $\pred'\,(\Suc\;n) = $ case $\pred'\;n$ of $(n_1,n_2) \Ra (n_1,\_)$, 
and then taking $\pred\;n$ to be the first component of $\pred'\;n$. 
(See also \cite[\S1.4.2]{pop-thesis}.) 

Initially, we had only developed the iteration theorems, but soon realized that 
several applications required full recursion (e.g., the complete development 
operator, the CPS transformation, etc.); and it was tedious to perform the above 
encoding over and over again. Consequently, we decided to formalize full recursion 
(for all three variants: substitution, swapping and substitution-swapping).  
We show here the full recursion variant of Theorem~\ref{th-rec}: 

A {\em full-recursion FS model} has the same components as an FS model, 
except that:
\begin{myitem}
\item $\OP$ takes term and abstraction inputs in addition to inputs from the model, i.e., 
has type $\opsym \ra \coll{(\indexx,\term)\;\inputt} \ra (\indexx,\T)\;\inputt 
\ra 
\coll{(\bindexx,\abs)\;\inputt} \allowbreak \ra 
(\bindexx,\A)\;\inputt \ra \T$
\item The freshness and substitution operators take additional term or abstraction arguments, 
e.g., $\FRESH : \varsort \ra \var \ra \coll{\term} \ra \T \ra \bool$
\end{myitem}

\begin{thm}\label{th-rec-full} \rm
The good terms and abstractions form the initial full-recursion FS model. Namely, 
for any full-recursion FS model as above, there exist the functions $\f:\term \ra \T$ 
and $\fAbs : \abs \ra \A$ that commute, on good terms, 
with the constructors and with substitution and preserve freshness in the same 
manner as in Theorem~\ref{th-rec}, mutatis mutandis. For example: 
\begin{myitem}
\item $\f\,(\Var\;\xs\;x) = \VAR\;\xs\;x$
\item $\f\,(\Op\;\delta\;\inp\;\binp) = 
\OP\;\delta\,\coll{\inp}\,(\lift\,\f\;\inp)\,\coll{\binp}\,(\lift\,\fAbs\;\binp)$
\item $\fresh\;\xs\;x\;X \,\Lra\, \FRESH\;\xs\;x\,\coll{X}\,(\f\;X)$
\end{myitem}
\end{thm}

\section{Applications of the Framework} \label{sec-app}

So far, we instantiated the binding signature to 
the syntaxes of the call-by-name and call-by-value
variants of the $\lambda$-calculus 
(the latter differing from the former by a separate syntactic category for values) 
and to that of many-sorted FOL---and applied it to results on these syntaxes. 
%

The first application was developed in 2010, when we proved strong normalization 
for System F \cite{pop-HOASOnFOAS}. The two employed syntaxes, of System F's Curry-style terms and types, are two copies 
of the $\lambda$-calculus syntax. 
The logical relation technique required in the proof made essential use of 
parallel substitution---and in fact was the incentive for us to go beyond unary 
substitution in the general theory. 
To streamline the development, on top of the first-order syntax we introduced 
HOAS-like definition and reasoning techniques, which were 
based on the general-purpose first-order ones shown in Section~\ref{sec-reas}. 

In subsequent 
work \cite{pop-recPrin}, we formalized several results about 
$\lambda$-calculus: 
the standardization and Church-Rosser theorems 
and the 
CPS translations between 
call-by-name and call-by-value calculi 
\cite{plotkin-CBNandCBVandLambda},  
an adequate HOAS representation of the calculus into itself, 
a sound interpretation via de Bruijn encodings \cite{cur-CategoricalCombinators}, and 
the isomorphism between different definitions of $\lambda$-terms: 
ours, the Nominal one \cite{UrbanTasson}, the 
locally named one \cite{pol-LocNamed2} and the Hybrid one \cite{momFelty-Hybrid4}. 
These results are centered 
around some translation/representation functions: CPS, HOAS, 
Church-Rosser complete development \cite{takahashi-CompleteDevelopment}, 
de Bruijn, etc.
---
these functions 
and their 
good properties (e.g., preservation of substitution, 
crucial in HOAS \cite{har-fra}) 
were obtained as 
instances of our recursion Theorem~\ref{th-rec}.   

Finally, in the context of certifying Sledgehammer's HOL to FOL encodings \cite{DBLP:journals/corr/BlanchetteB0S16}, 
we formalized fundamental results in many-sorted FOL \cite{blanchette-frocos2013}, 
including the completeness, compactness, 
Skolemization and downward L\"{o}wenheim-Skolem theorems. Besides the ubiquitous 
employment of the properties of freshness of substitution from Section~\ref{subsec-termsTh}, 
we 
used the interpretation Theorem~\ref{th-sem} for the FOL semantics 
and the recursion Theorem~\ref{th-rec} for bootstraping quickly the (technically quite tricky) Skolemization 
function.